\documentclass[preprint,3p,times,twocolumn]{elsarticle}

\usepackage{graphicx}

\usepackage{amssymb}

\usepackage{lineno}

\usepackage{tabularx} 

\usepackage{newtxmath, newtxtext} 

\usepackage{amsmath}
\usepackage{bm}  

\begin{document}

\begin{frontmatter}

\title{EuPRAXIA@SPARC\_LAB\\ Design study towards a compact FEL facility at LNF}



\author[lnf]{M. Ferrario \corref{cor1}}
\ead{massimo.ferrario@lnf.infn.it}
\cortext[cor1]{Corresponding author}

\author[lnf]{D. Alesini}
\author[lnf]{ M. P. Anania}
\author[enea]{M. Artioli}
\author[infnmi]{ A. Bacci}
\author[uniss]{S. Bartocci}
\author[lnf]{R. Bedogni}
\author[lnf]{ M. Bellaveglia}
\author[lnf]{A. Biagioni}
\author[lnf]{ F. Bisesto}
\author[ililcnr]{F. Brandi}
% ...


\address[lnf]{Laboratori Nazionali di Frascati - INFN, via E. Fermi 40, 00044 Frascati, Italy}
\address[infnmi]{INFN sect. Milano, Via Celoria 16, 20133 Milan, Italy}
\address[infntri]{INFN sect. Trieste, Via Valerio 2, 34127 Trieste, Italy}
\address[lns]{Laboratori Nazionali del Sud - INFN, via S.Sofia 62, 95123 Catania, Italy}
\address[eneafr]{ENEA - Centro Ricerche Frascati, Via E. Fermi 45, 00044 Frascati, Italy}
\address[eneabo]{ENEA  -Centro Ricerche Bologna, Via Martiri Monte Sole 4, 40129 Bologna, Italy}
\address[ililcnr]{Intense Laser Irradiation Laboratory (ILIL), Istituto Nazionale di Ottica (INO), Consiglio Nazionale delle Ricerche (CNR), Via G. Moruzzi 1, 56124 Pisa, Italy
2 and INFN sect. Pisa, Largo Pontecorvo 3, 56127 Pisa, Italy}
\address[cern]{CERN, CH-1211 Geneva 23, Switzerland}
\address[unitv]{Universit\'{a} degli Studi di Roma Tor Vergata and INFN sect., Via della Ricerca Scientifica 1, 00133 Rome, Italy}
\address[unimi]{Universit\'{a} degli Studi di Milano and INFN sect., Via Celoria 16, 20133 Milan, Italy}
\address[unisap]{Universit\'{a} degli Studi di Roma La Sapienza and INFN sect., P.le Aldo Moro 2, 00185 Rome, Italy}
\address[uniss]{Universit\'{a} degli Studi di Sassari, Dip. di Architettura, Design e Urbanistica ad Alghero, Palazzo del Pou Salit - Piazza Duomo 6, 07041 Alghero, Italy}
\address[ism]{ISM-CNR, Basovizza Area Science Park, Elettra Lab, 34149 Trieste - Italy}
\address[elettra]{Elettra-Sincrotrone Trieste, Area Science Park, 34149 Trieste, Italy}
\address[ricmass]{RICMASS, Rome International Center for Materials Science Superstripes, 00185 Rome, Italy}
\address[unij]{Racah Institute of Physics, The Hebrew University of Jerusalem, 91904 Jerusalem, Israel}
\address[ucla]{Department of Physics and Astronomy, University of California Los Angeles, Los Angeles, California 90095, USA}


\begin{abstract}
On the wake of the results obtained so far at the SPARC\_LAB test-facility at the Laboratori Nazionali di Frascati (Italy), we are currently investigating the possibility to design and build a new multi-disciplinary user-facility, equipped with a soft X-ray Free Electron Laser (FEL) driven by a $\sim$1 GeV high brightness linac based on plasma accelerator modules. This design study is performed in synergy with the EuPRAXIA design study. In this paper we report about the recent progresses in the on going design study of the new facility.
\end{abstract}


\end{frontmatter}



\section{Introduction}
\label{intro}

It is widely accepted by the international scientific community that a fundamental milestone towards the realization of a plasma driven future Linear Collider will be the integration of a high gradient accelerating plasma modules in a short wavelength Free Electron Laser (FEL) user facility \cite{Assmann}. The capability of producing the required high quality beams and the operational reliability of the plasma accelerator modules will be certainly certified when such an advanced radiation source will be able to drive external user experiments. It is further expected that there will be unique photon-beam characteristics that give notable advantages to such a plasma based light source. These include enabling ultra-short photon pulses based on high brightness electron beams that break the attosecond barrier and, when used in combination with next generation undulators, shorter wavelength photons at notably lower electron beam energy. The realization of such a light source will serve as a required stepping stone for High Energy Physics (HEP) applications and will be a  new tool for photon science in its own right.
The Horizon2020 Design Study EuPRAXIA (European Plasma Research Accelerator with eXcellence In Applications) \cite{Walker11} will in October 2019 propose a first European Research Infrastructure that is dedicated to demonstrate explotation of plasma accelerators for users. Developing a consistent set of beam parameters produced by a plasma accelerator able to drive a short wavelength FEL is one of the major commitments of the EuPRAXIA Design Study. 
At present, five different EuPRAXIA configurations are under investigation \cite{Walker11}, based on a laser and/or a beam driven plasma acceleration approaches. The first iteration of the design parameter goals was defined in October 2016 \cite{Walker}. The first EuPRAXIA phase will target at a Self Amplified Spontaneous Emission (SASE) FEL design driven by a 1 GeV electron beam thus producing radiation in the soft x-ray range as a proof of principle demonstration. The 1 GeV operation aims at a very compact version of soft X-ray FEL's like FLASH \cite{Ayvazyan} in Hamburg or FERMI \cite{fermi} in Trieste. This would be a first breakthrough offering interesting light pulses for first pilot users. The second, more demanding phase, will lead the way for a plasma-based SASE-FEL design at 5 GeV electron beam energy in the hard X-ray range. Practical realization of such an FEL however necessarily requires the experience from a previous iteration. 
The site selection for EuPRAXIA will be performed during the Preparatory Phase (expected in the years 2020-2022), following the delivery of the Conceptual Design Report (2019) and the inclusion of EuPRAXIA into the European Strategy Forum on Research Infrastructures (ESFRI) roadmap (expected in 2020). 

\begin{figure*}[hbt]
\includegraphics[width=\textwidth]{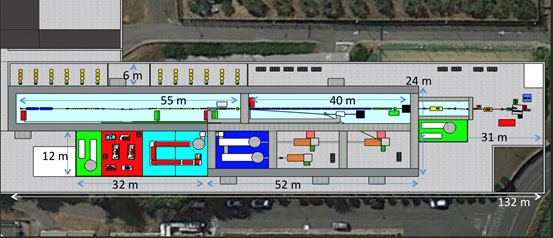}
\caption{The layout of the EuPRAXIA@SPARC\_LAB infrastructure}
\label{fig:eusparc}
\end{figure*}

\section{The EuPRAXIA@SPARC\_LAB concept}
In this paper we discuss the EuPRAXIA@SPARC\_LAB project, intended to put forward the Laboratori Nazionali di Frascati (LNF) in Italy as host of the EuPRAXIA European Facility. The new infrastructure will be able to accommodate any machine configuration resulting from the EuPRAXIA Design Study that will find within the new LNF infrastructure the necessary technological background. 
In order to achieve this goal and to meet the EuPRAXIA requirements, some important preparatory actions must be taken at LNF:
\begin{itemize}
\item provide LNF with a new infrastructure with the size of about 130 m x 30 m, as the one required to host the EuPRAXIA facility;
\item design and build the first-ever 1 GeV X-band RF linac and an upgraded FLAME laser up to the 0.5 PW range;
\item design and build a compact FEL source, equipped with user beam line at 4--2 nm wavelength, driven by a high gradient plasma accelerator module.
\end{itemize}
The EuPRAXIA@SPARC\_LAB facility by itself will equip LNF with a unique combination of a high brightness GeV-range electron beam generated in a state-of-the-art linac, and a 0.5 PW-class laser system. Even in the case of LNF not being selected and/or of a failure of plasma acceleration technology, the infrastructure will be of top-class quality, user-oriented and at the forefront of new acceleration technologies. These unique features will enable at LNF new promising synergies between fundamental physics oriented research and high social impact applications, especially in the domain of Key Enabling Technologies (KET) and Smart Specialisation Strategies (S3), as supported by EU research funding programs. EuPRAXIA@SPARC\_LAB is in fact conceived by itself as an innovative and evolutionary tool for multi-disciplinary investigations in a wide field of scientific, technological and industrial applications. It could be \emph{progressively} extended to be a high brightness "particles and photons beams factory": it will be eventually able to produce electrons, photons (from THz to $\gamma$-rays), neutrons, protons and positrons, that will be available for a wide national and international scientific community interested to take profit of advanced particle and radiation sources.

The EuPRAXIA@SPARC\_LAB project requires the construction of a new building to host the linac, the FEL, the experimental room and the support laboratories. The new facility will cover approximately an area of 4000 m$^2$. The layout of the EuPRAXIA@SPARC\_LAB infrastructure is schematically shown in Figure \ref{fig:eusparc}.

From left to right one can see a 55 m long tunnel hosting a high brightness 150 MeV S-band RF photoinjector equipped with a hybrid compressor scheme based on both velocity bunching \cite{Serafini, Ferrario7, Giribono} and magnetic chicane. The energy boost from 150 MeV up to a maximum 1 GeV will be provided by a chain of high gradient X-band RF cavities \cite{Diomede, Vaccarezza}. At the linac exit a 5 m long plasma accelerator section will be installed, which includes the plasma module ($\sim$0.5 m long) \cite{Filippi11} and the required matching \cite{Chiadroni, Brentegani} and diagnostics sections \cite{Cianchi, Marongiu}. In the downstream tunnel a 40 m long undulator hall is shown, where the undulator chain will be installed \cite{Petrillo}. Further downstream after a 12 m long photon diagnostic section \cite{Villa} the users hall is shown \cite{notaextrim}. Additional radiation sources as THz and $\gamma$-ray Compton sources are foreseen in the other shown beam lines. The upper room is dedicated to Klystrons and Modulators. In the lower light-blue room will be installed the existing 300 TW FLAME laser \cite{Bisesto} eventually upgraded up to 500 TW. The plasma accelerator module can be driven in this layout either by an electron bunch driver (PWFA scheme) \cite{Marocchino} or by the FLAME laser itself (LWFA scheme) \cite{Rossi}. A staged configuration of both PWFA and LWFA schemes will be also possible in order to boost the final beam energy beyond 5 GeV. In addition FLAME is supposed to drive plasma targets in the dark-blue room in order to drive electron and secondary particle sources that will be available to users in the downstream 30 m long user area. 

\begin{figure}[hbt]
\includegraphics[width=0.475\textwidth]{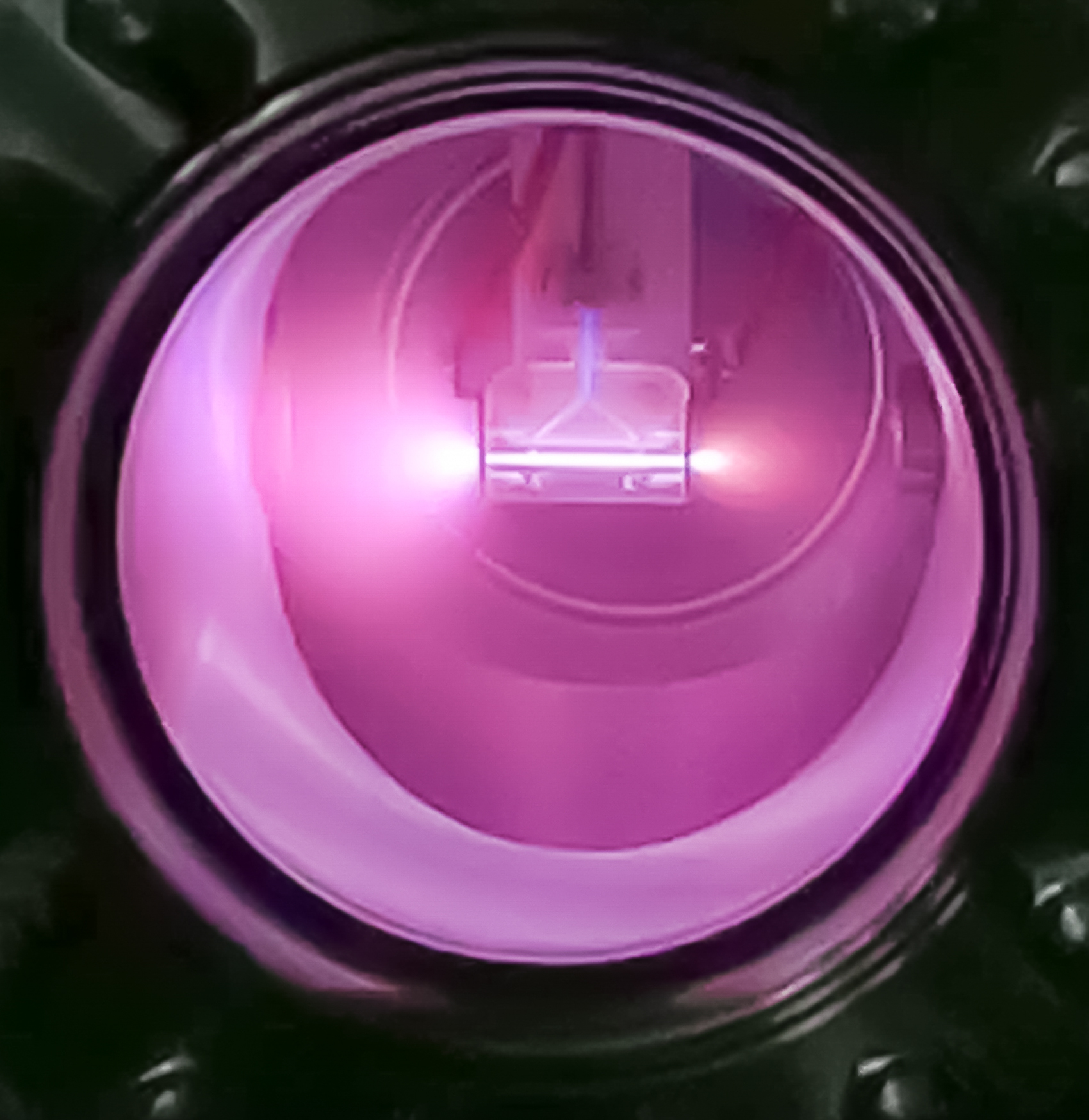}
\caption{Discharge capillary installed in the SPARC\_LAB beamline}
\label{fig:capillare}
\end{figure}

One of the most innovative device of the project is the plasma accelerating module, in one of its possible configurations 
\cite{Zigler,Brill,Zigler2}, see Figure \ref{fig:capillare}. 

It consists in a 10 cm long, 0.5 mm diameter capillary tube \cite{Filippi} in which the plasma is produced by a high voltage discharge in hydrogen. Another fundamental component included in our proposal is the X-band accelerating technology adopted for the 1 GeV RF drive linac \cite{Wuensch}. It is a very interesting option because it allows to reduce the overall drive linac length, taking profit of the high gradient (up 80 MV/m \cite{Diomede}) operation of the X-band accelerating structures. In addition it will allow implementing at LNF in the next 2 years the state of the art high gradient RF technology. This technology has already shown its usefulness for medical and industrial applications but it is also another possible technological option for compact radiation sources and for the future Linear Collider \cite{HG2017}. 

In the PWFA scenario driven by a single electron bunch, the peak accelerating field is, in principle, limited to twice the value o the peak decelerating field within the bunch (transformer ratio R=2). Therefore the maximum possible energy gain for a trailing bunch is less than twice the incoming driver energy. In this regime a driver bunch energy of 600 MeV is enough to accelerate the witness bunch up to 1 GeV. A method to increase the energy gain is the so called \emph{ramped bunch train} \cite{Schutt} and consists of using a train of $N_T$ equidistant bunches, wherein the charge increases along the train producing an accelerating field resulting in a higher transformer ratio. For this application, it is essential to create trains of high-brightness tens of femtosecond long microbunches with stable and adjustable length, charge and spacing. A lot of efforts are now ongoing worldwide to produce the required bunch train configurations \cite{Muggli}. The method we will use to achieve the required bunch train quality is based on the \emph{Laser Comb Technique} \cite{Ferrario} that has been tested with the SPARC\_LAB photoinjector \cite{ferrariocomb}. Higher witness bunch energy will thus be accessible when the Comb technique will be implemented. With a transformer ratio > 4 the 5 GeV threshold will be achievable with a 1 GeV driver bunch energy, thus exploiting the full energy provided by the X-band linac. Other methods pushing R, apart from bunch trains, are possible as discussed in \cite{Shaped} and they will be also considered.

In the LWFA scenario the 0.5 PW upgrade of the FLAME laser \cite{Bisesto} is a necessary step to keep the FLAME laser in the group of leading installations and further establish expertise on advanced laser sciences. High energy staging in combination with high brightness beam external injection will be the main application of the upgraded FLAME system, leading to multi-GeV high brightness electron beam production as required by the final EuPRAXIA goals. 

We have investigated the possibility to fulfill the 1 GeV EuPRAXIA parameters \cite{Walker}. To support our design in both plasma acceleration options (PWFA and LWFA) we have performed  Start To End Simulations with promising results, as discussed in \cite{Giribono,Vaccarezza,Petrillo, Marocchino, Rossi}. In Table \ref{eusparcpar} the achieved parameters are reported. The reported performances show that our FEL design, driven by a plasma accelerator in SASE configuration, is expected to meet the challenging requests for the new generation synchrotron radiation sources. We have investigated also the possibility to drive the FEL with higher charge/bunch i.e. up to 200 pC, in order to produce a larger number of photons as required by some application.  This is possible in a conventional configuration without using the plasma module and the results of Start-To-End simuations for this case are also shown in Table \ref{eusparcpar}. Possible applications of the EuPRAXIA@SPARC\_LAB FEL source at 3 nm are described in the next paragraph.

\begin{table*}[htb]
\centering
\begin{tabularx}{\textwidth}{| >{\raggedright}p{3cm} | >{\centering}p{1cm} | >{\centering}X | >{\centering}X | >{\centering}X|} 
\hline 
 & \textbf{Units}  & \textbf{1 GeV PWFA with Undulator Tapering}  & \textbf{1 GeV LWFA with Undulator Tapering} & \textbf{1 GeV  with X-band linac only 200 pC }  \tabularnewline
\hline
\textbf{Bunch charge}  & pC & 29 & 26.5 &  200 \tabularnewline
\hline
\textbf{Bunch length rms}  & fs & 11.5 & 8.4 &  55.6 \tabularnewline
\hline
\textbf{Peak current}  & kA & 2.6 & 3.15 &  1.788 \tabularnewline
\hline
\textbf{Rep. rate}  & Hz & 10 & 10 &  10 \tabularnewline
\hline
\textbf{Rms Energy Spread}  & \% & 0.73 & 0.81 &  0.05 \tabularnewline
\hline
\textbf{Slice Energy Spread}  & \% & 0.022 & 0.015 &  0.02 \tabularnewline
\hline
\textbf{Average rms norm. emittance}  & $\mu$m & 0.6 & 0.47 &  0.5 \tabularnewline
\hline
\textbf{Slice norm. emittance}  & $\mu$m & 0.39 -- 0.309 & 0.47 &  0.4 -- 0.37 \tabularnewline
\hline
\textbf{Slice Length}  & $\mu$m & 1.39 & 1.34 &  1.66 \tabularnewline
\hline
\textbf{Radiation wavelength}  & nm & 2.79 & 2.7 &  2.87 \tabularnewline
\hline
$\boldsymbol{\rho}$  & x 10$^{-3}$ & 2 & 2 &  1.55 (1.38) \tabularnewline
\hline
\textbf{Undulator period}  & cm & 1.5 & 1.5 &  1.5 \tabularnewline
\hline
\textbf{K}  &  & 0.987 & 1.13 &  0.987 \tabularnewline
\hline
\textbf{Undulator length}  & m & 30 & 30 &  16 -- 30 \tabularnewline
\hline
\textbf{Saturation power}  & GW & 0.850 -- 1.2 & 1.3 &  0.12 -- 0.33 \tabularnewline
\hline
\textbf{Energy}  & $\mu$J & 63 & 63.5 &  64 -- 177 \tabularnewline
\hline
\textbf{Photons/pulse}  & x 10$^{11}$ & 8.8  & 8.6  &  9.3 -- 25.5  \tabularnewline
\hline
\textbf{Bandwidth}  & \% & 0.35 & 0.42 &  0.24 -- 0.46 \tabularnewline
\hline
\textbf{Divergence}  & $\mu$rad & 49 & 56 &  28 -- 27 \tabularnewline
\hline
\textbf{Rad. size}  & $\mu$m & 210 & 160 &  120 -- 200 \tabularnewline
\hline
\end{tabularx}
\caption{Beam parameters for plasma and conventional RF linac driven FEL.}
\label{eusparcpar}
\end{table*}

\section{The EuPRAXIA@SPARC\_LAB scientific goals}

The experimental activity will be initially focused on the realization of a plasma driven short wavelength FEL with one user beam line, according to the beam parameter reported in the Table \ref{eusparcpar}. This goal is already quite challenging but it is affordable by the EuPRAXIA@SPARC\_LAB collaboration and will provide an interesting FEL radiation spectrum in the so called "water window". The first foreseen FEL operational mode is based on the Self Amplification of Spontaneous Radiation (SASE) mechanism \cite{Bonifacio} with tapered undulators. More advanced schemes like Seeded and Higher Harmonic Generation configurations will be also investigated. 
The users endstation, called EX-TRIM (Eupraxia X-ray Time Resolved coherent IMaging), will be designed and built to allow performing a wide class of experiments using the schematic apparatus discussed in \cite{notaextrim,Chiadroni}. As specific example of EuPRAXIA@SPARC\_LAB applications it is worth remarking that the FEL radiation in the soft X-ray spectrum open possibilities for novel imaging methodologies and time-resolved studies in material science, biology and medicine, along with non-linear optics applications, for example:

\emph{Coherent Imaging of Biological samples in the water window} Exploiting the coherence of the FEL beam and its wavelength falling within the "water window", 2D and 3D images of biological samples in a wet environment can be obtained with high contrast with respect to the surrounding medium. This means that a wide class of biological objects, including protein clusters, viruses and cells can be profitably studied.

\emph{Clusters and nanoparticles} In particular, great interest arises in the correlations between the geometric structure and electronic properties of variable size clusters, underlying changes in optical, magnetic, chemical and thermodynamic properties. In the spectral range from 3 to 5 nm envisaged for the FEL source, physical processes involving core levels are important. 

\emph{Laser ablation plasma} Laser ablation/desorption techniques are utilized extensively across a diverse range of disciplines, including production of new materials, and both extrinsic and in situ chemical analysis. In the case of ablation the use of ultra-fast laser pulses provides a powerful means of machining a wide variety of materials, including biological tissue. The absence of thermal relaxation of the energy allows unprecedented precision and essentially no associated damage, a fact that has stimulated considerable interest also for industrial processes and applications. Electronically induced surface reactions in semiconductors, metal/adsorbate systems and multiphase composite materials could be investigated.

\emph{Condensed Matter Science}  Coherent Diffraction Imaging (CDI) experiments tackling many open questions in Condensed Matter physics. For instance, the quest for smaller and faster magnetic storage units is still a challenge of the magnetism. The possibility to study the evolution of magnetic domains with nanometer/femtosecond spatial/temporal resolution will shed light on the elementary magnetization dynamics such as spin-flip processes and their coupling to the electronic system. 

\emph{Pump and probe experiments} As an example, resonant experiments with pulses tuned across electronic excitation will open up the way towards stimulated Raman or four wave mixing spectroscopies.

In addition, the upgrade of the FLAME laser system to 0.5 PW power and dual beam capability will enable new regimes of plasma-based particle accelerators and will enable to access the region of high electromagnetic fields of non-linear and quantum electrodynamics (QED) where new fundamental physics processes and promising new radiation emission mechanisms can be explored.  The science cases that will be developed with the FLAME 0.5 PW upgrade includes: 

\begin{itemize}
\item Electron acceleration beyond the GeV (including external injection and/or Trojan horse scheme , high energy staging, etc.);
\item QED and generation of high energy radiation;
\item Proton and ion acceleration beyond the TNSA regime;
\item Betatron and Compton scattering radiation sources;
\item Compact Neutron source \cite{neutron}.
\end{itemize}

\section{Conclusions}
\label{conc}

 Together with the driving motivation to candidate LNF to host the EuPRAXIA facility, the realization of the EuPRAXIA@SPARC\_LAB infrastructure at the LNF by itself will allow INFN to consolidate a strong scientific, technological and industrial role in a competing international context. A national multi-purpose facility, along the scientific applications discussed in the following sections, not only paves the road for a strong role for the Italian contribution to the European EuPRAXIA one, but also to possible future large HEP international projects. We are confident that this project will represent a further step forward in the mainstream of a long lasting history of success in particle accelerators development in Frascati. 

\section*{Acknowledgments}
This work was supported by the European Union's Horizon 2020 research and innovation programme under grant agreement No. 653782.





\section*{Bibliography}
\bibliographystyle{elsarticle-num}
\bibliography{bib_proc_eaac17.bib}








\end{document}